\documentclass{article}
\usepackage[dvips,final]{graphics}

\newcommand{\bra}[1]{\left\langle #1 \right|}
\newcommand{\ket}[1]{\left| #1 \right\rangle}

\begin{document}

\title{Selecting molecules in the vibrational and rotational ground state by
  deflection}

\author{P. Domokos\thanks{Present address: Institute f\"ur Theoretische
    Physik, Universit\"at Innsbruck, Technikerstrasse 25, A-6020 Innsbruck,
    Austria}, T. Kiss\thanks{Also at: University of
    P\'ecs, Ifj\'us\'ag u. 6. H-7624 P\'ecs, Hungary}, J. Janszky\\
\
\\
Research Institute for Solid State Physics and Optics,\\ H-1525 Budapest,
  P.~O.~Box 49, Hungary}

\date{December 15, 2000}

\maketitle

\begin{abstract}
  A beam of diatomic molecules scattered off a standing wave laser mode splits
  according to the rovibrational quantum state of the molecules.  Our
  numerical calculation shows that single state resolution can be achieved by
  properly tuned, monochromatic light. The proposed scheme allows for
  selecting non-vibrating and non-rotating molecules from a thermal beam,
  implementing a laser Maxwell's demon to prepare a rovibrationally cold
  molecular ensemble.
\end{abstract}


\newpage
Manipulating the motion of massive particles by using the mechanical effect of
light has been a rapidly developing field since the discovery of lasers.
Understanding the fine details of the light--matter interaction on a
microscopic level contributed much to the advent of impressive applications,
including laser cooling and trapping of atoms and atom optics~\cite{nobel}.
The adaptation of these techniques for molecules would undoubtedly open new
perspectives. Recently, pioneering experiments have been reported in this
direction, for trapping \cite{takekoshi98,doyle} and decelerating
\cite{bethlem99} molecules or for interferometry \cite{chapman95}, and also
for fundamental tests of quantum mechanics \cite{arndt99} with molecules. It
is, however, nontrivial to extend the optical methods to molecules because of
the complexity of their energy level structure. Optical cooling of molecules
\cite{bahns00} would require new approaches, for example, such as coupling the
particle to a far-detuned light field inside an optical cavity
\cite{vuletic00}.

The essential difference stems from the additional mechanical degrees
of freedom. The vibrational and rotational modes lead to densely
spaced levels.  Typical excitations are orders of magnitude smaller
than the electronic transitions. A usual thermal beam consists then of
molecules in various rovibrational quantum states. Cooling
translation, vibration, and rotation simultaneously is a difficult
problem; a complex scheme with multiple laser beams has been proposed
in \cite{bahns96}.  However, there is another option to obtain molecules
with lower energy: to sort the thermal ensemble by means of
state-sensitive interactions. A Stern-Gerlach-type experiment with
inhomogeneous microwave field has been first performed to separate the
rotational states constituting the resonant transition within a given
vibrational band \cite{hill75}. Rotational state dispersion has been
shown to be present in various molecular optics settings using
far-detuned laser fields \cite{seideman97a}. In this paper we show
that a Maxwell demon like operation can be achieved in a different
regime of the molecule-light interaction. That is, a single
rovibrational state, possibly the ground state, can be filtered from a
thermal beam. Owing to the high degree of control facilities, the
proposed scheme can be applied to a quite general class of molecular
states.

Deflection of a CS$_2$ and I$_2$ beam using the nonresonant dipole
force has recently been demonstrated \cite{corkum}. The laser was well
detuned from a given electronic transition and all the molecules
experienced very nearly the same nonresonant dipole force regardless
of what their internal quantum state was.  In order to accommodate the
nonresonance condition for all the transitions, detuning on the scale
of the electronic transition frequency was applied.  More
sophisticated control of the mechanical effect of light could be
obtained with a laser whose bandwidth is comparable with the linewidth
of the rovibrational transitions in the spectrum. With detunings on
the order of the rotational level-spacing, less laser power is needed
to have the same mechanical effect.

The reaction of atoms and molecules to radiation is usually described
by the frequency--dependent polarizability tensor. This is a good
approximation for a transparent medium, when the radiation frequency is
far away from any resonances in the matter. The polarizability can
then be calculated from the quantum mechanical mean value of the
dipole operator, usually by employing perturbation theory. This
approach is sufficient to interpret the deflection experiment
\cite{corkum}, and many other schemes for manipulating molecular motion
\cite{karczmarek99,friedrich95}. However, the simple concept of a
state-independent polarizability does not hold when the monochromatic
laser frequency is close to some individual rovibronic resonances. In
this case one must return to a more fundamental quantum mechanical
treatment relying on the dipole operator itself. Otherwise the
standard theoretical framework for treating molecular optics
\cite{seideman}, 
that is the adiabatic separation of the rovibrational modes and the
center-of-mass motion can be used.

\begin{figure}
\resizebox{\textwidth}{!}{%
  \includegraphics*{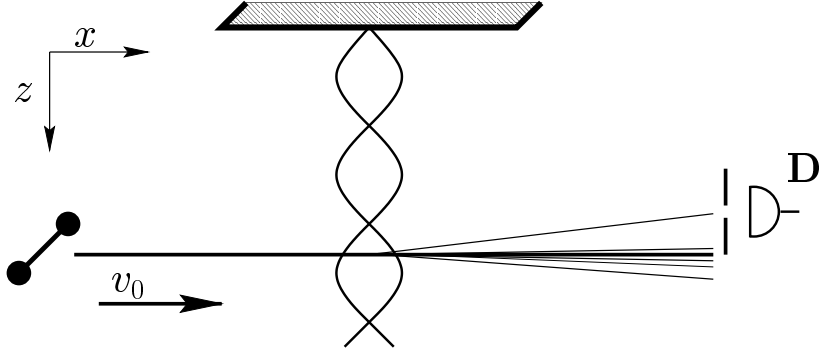}
} 
\caption{Scheme of deflecting molecules quantum state selectively.
      A beam of diatomic molecules with velocity $v_0$ is scattered off a
      monochromatic, standing wave laser mode. 
      The detector {\bf D} measures the
      deflected beam intensity along the $z$ axis.}
    \label{fig:scheme}
\end{figure}
We consider a beam of diatomic molecules (see Fig.~1) moving with
velocity $v_0$ in the $x$ direction and crossing at a right angle a CW
standing-wave laser mode. The interaction energy, in the dipole
approximation, reads $V=-{\hat {\bf d}} {\bf E}$. The electric field
is linearly polarized, and its spatial variation is given by ${\bf
E}={\bf E}_0\,
\exp{[-(x^2+y^2)/2w^2]}\, \cos{k z} \cos{\omega t}$. The molecules experience
a force proportional to the gradient of the inhomogeneous interaction
energy.  Due to this light force, the molecules accumulate transverse
momentum, $M v_z$. However, the displacement in this direction is
negligible during the interaction (Raman--Nath regime). The
laser-induced potential is assumed to be much smaller than the kinetic
energy; hence the longitudinal velocity $v_x
\equiv v_0$ remains constant during the whole passage.

The spectrum of the diatomic molecule in the Hund's case $a$ is
defined  \cite{kovacs} by the Hamiltonian
\begin{equation}
  \label{H_0}
  H_0=\sum E(n,\nu,J,M,\Omega) \ket{n \nu J M \Lambda \Sigma}\bra{n \nu J M
  \Lambda \Sigma}\; ,
\end{equation}
where $n$ labels the electronic states, $\nu$ is the vibrational, $J$
and $M$ are the usual rotational quantum numbers, $\Lambda$ and
$\Sigma$ are the projections of the total electronic angular momentum
and spin on the intermolecular axis, respectively, and $\Omega$ is
their sum. Nuclear spin is ignored; however, we will comment on this
point later on. Within the Born--Oppenheimer approximation, the
electronic and nuclear wave functions can be separated as follows
\begin{equation}
  \label{states}
  \ket{n \nu J M \Lambda \Sigma}=\Phi_{n \Lambda \Sigma}(\tau_e,r) \,
  R_{n \nu}(r) \, u_{J M \Omega}(\theta, \phi)\; ,
\end{equation}
where the variable $\tau_e$ incorporates all the electron coordinates,
$r$ is the internuclear distance, and $\theta$ and $\phi$ define the
orientation of the molecule. The electronic spin part of the quantum
state is omitted. The energies for low $\nu$ and $J$ are usually
approximately given by the expression
\begin{eqnarray}
  \label{energies}
  E(&n&,\nu,J,M,\Omega)= \nonumber\\
          &&E_{\mbox{\tiny el}}(n)+\hbar\omega_e(\nu+1/2)-
\hbar\omega_e x_e (\nu+1/2)^2\nonumber\\ 
          &&+B_{\nu}[J(J+1)-\Omega^2]-D_\nu[J(J+1)-\Omega^2]^2\; ,
\end{eqnarray}
where the first term is the electronic energy, the next two terms describe the
anharmonic vibration, and the last two correspond to the energy of a symmetric
top (the standard notation \cite{kovacs,herzberg} is used).

Let us suppose that only the fundamental $f$ and one excited $e$ electronic
states are effectively coupled by the electromagnetic mode. The interaction
Hamiltonian can be restricted to
\begin{eqnarray}
  \label{H_total}
  V &=& - {\bf E}_0 e^{-(x^2+y^2)/2w^2}\cos{kz}\cos{\omega t}\nonumber\\
  && \cdot \sum_{\nu J M} \sum_{\nu' J' M'} 
      {\bf d}_{f \nu J M}^{e \nu' J' M'} 
  \ket{f \nu J M}\bra{e \nu' J' M'} + \mbox{h.~c.} \; 
\end{eqnarray}
The electric field ${\bf E}_0$ is linearly polarized in an arbitrary
direction perpendicular to the mode axis $z$. It is enough to consider
the dipole moment in the direction of the polarization. We resort to
the Franck--Condon approximation to calculate the matrix elements of
the dipole operator, i.e., the induced electric dipole moment is supposed to be
independent of the internuclear distance. This separation is not
fundamental, but only convenient for purposes of discussion. The
matrix element can then be decomposed into three factors, $d_{f \nu J
M}^{e \nu' J' M'}= d_{f}^{e} R_{\nu}^{\nu'} L_{J M \Omega}^{J'
M' \Omega'}$. The electric dipole is given by the integral
\begin{equation}
  \label{d12}
  d_{f}^{e}=\overline{\int \Phi_e^*(\tau_e,r)  {\hat d} \Phi_f(\tau_e,r)
  d\tau_e}\; ,
\end{equation}
averaged over the nuclear distances $r$. 
The next term,
\begin{equation}
  \label{FCdef}
  R_{\nu}^{\nu'} = \int R_{e \nu'}^* R_{f \nu} r^2 dr\; ,
\end{equation}
is the vibrational overlap factor (the square root of the
Franck--Condon factor) that we will calculate using Morse-potential
eigenfunctions. Finally, a rotation-dependent factor follows which can
be analytically evaluated for the symmetric top eigenfunctions.  Two
cases can be distinguished:
\begin{equation}
L_{J M \Omega}^{J' M' \Omega'}= \int u_{J' M' \Omega'}^* u_{J M \Omega}^{}\, 
  \left\{ \begin{array}{c} \cos{\theta}\\ \sin{\theta} \end{array}\right\}\, 
  \sin{\theta} d\theta d\phi 
   \; ,
\end{equation}
where $\cos{\theta}$ is used for transitions with $\Delta\Omega=0$, and
$\sin{\theta}$ is used for transitions with $\Delta\Omega
= \pm 1$. 

The Hamiltonian, in the rotating wave approximation, can be transformed to the
form 
\begin{eqnarray}
  \label{Htotal}
  &&{\hat H}_{\mbox{\tiny eff}}=\sum_{\nu J M} \sum_{\nu' J' M'}\nonumber\\ 
  &&\bar E_{\nu J M}^{\nu' J' M'}
  \left(\ket{f \nu J M}\bra{f \nu J M} +
      \ket{e \nu' J' M'}\bra{e \nu' J' M'}\right) \nonumber\\
  &&+\frac{1}{2}\hbar\delta_{\nu J M}^{\nu' J' M'} 
   \left(\ket{e \nu' J' M'}\bra{e \nu' J' M'} - 
        \ket{f \nu J M}\bra{f \nu J M}\right) \nonumber \\ 
  &&+\hbar g_{\nu J M}^{\nu' J' M'} f({\bf R}) 
  \left(\ket{f \nu J M}\bra{e \nu' J' M'} + h.~c.\right)\; .
\end{eqnarray}
The first term defines the energy zero level for each rovibronic
transition $\ket{f \nu J M} \leftrightarrow \ket{e \nu' J' M'}$. The
second term is the corresponding detuning from the laser frequency,
$\hbar\delta_{\nu J M}^{\nu' J' M'} \equiv E(e \nu' J' M') - E(f \nu J
M) -\hbar\omega$. The coupling strength is given by $g_{\nu J M}^{\nu' J'
M'} \equiv {\bf d}_{f \nu J M}^{e \nu' J' M'} {\bf E}_0/\hbar$.  The
mode function $f({\bf R})$ collects the spatially varying factors. The
coupling constant can be expressed in terms of measurable quantities
as
\begin{equation}
  \label{coupling}
  \left(g_{\nu J M}^{\nu' J' M'}\right)^2 = 
  \frac{3 (\lambda_{f \nu J}^{e \nu' J'})^3}{16 \pi^2 \hbar c}
  \Gamma_{f \nu J}^{e \nu' J'} \frac{(2 J'+1) |L_{J M \Omega}^{J' M'
  \Omega'}|^2}{S(J,J')} \frac{I}{A}\; ,
\end{equation}
where $\Gamma_{f \nu J}^{e \nu' J'}$ is the natural linewidth, and $S(J,J')$
is the H\"onl-London factor. The laser parameters are the intensity $I$ in
units of photons/second and the effective beam area $A$.

The fine tuning of the laser is the crucial point of our scheme.  The laser
induces nonresonant interactions if the condition $(\delta_{\nu J M}^{\nu' J'
  M'})^2 \gg (g_{\nu J M}^{\nu' J' M'})^2 $ holds for each transition. The
laser is detuned from all the transitions, $\delta_{\nu J M}^{\nu' J' M'} >
\Gamma_{\nu J M}^{\nu' J' M'}$, and the smallest detunings are on the order of
the rotational level spacing $B_\nu$.  The nonresonance condition on this
scale limits the coupling $g_{\nu J M}^{\nu' J' M'} \ll B_\nu$, and thus the
laser intensity.

The molecules are initially in the electronic ground state, and the
vibrational and rotational states are populated according to a thermal
distribution. The Hamiltonian (\ref{Htotal}) describes the
laser-induced dynamics of the molecule except for the translational
motion.  Interaction with the vacuum modes yielding spontaneous
emission was omitted, since the upper electronic states are not
populated by the nonresonant excitation.  The translational motion is
treated classically in our model. The center-of-mass (CM) coordinate
appears as a parameter in $\hat H_{\mbox{\tiny eff}}$ via the mode
function $f({\bf R})$. Motion in the $y$ direction is irrelevant with
respect to the deflection scheme. We insert $v_0 t$ in the coordinate
$x$, thus making the problem explicitly time-dependent. For slow
enough motion, the system evolves adiabatically. In the nonresonant
limit the eigenstates of $\hat H_{\mbox{\tiny eff}}$, which can be
viewed as dressed rovibronic states, are close to those of the
unperturbed ${\hat H}_0$.  Transitions can be neglected between these
eigenstates. The eigenenergies are shifted (AC Stark effect) by an
amount that can be well approximated by taking only the most
significant transition into account for every rovibronic state. In
fact, this approximation is equivalent to the decomposition of
$H_{\mbox{\tiny eff}}$ into 2 by 2 blocks. Corrections can be
generated systematically by extending the block size with the
inclusion of weaker and weaker couplings.  The consistency of this
method is based on the hierarchy among the couplings that can be set
up according to the $g_{\nu J M}^{\nu' J' M'}/\delta_{\nu J M}^{\nu'
J' M'}$ ratios. This is not a perturbational expansion: whenever an
additional coupling term in $\hat H_{\mbox{\tiny eff}}$ is
considered, its contribution is completely included.  In a first
approximation, the energy associated with the state $\ket{f \nu J M}$
reads
\begin{eqnarray}
  \label{eigenenergies}
 && E_{\mbox{\tiny dress}}(f \nu J M) =\nonumber\\ 
 &&\bar E_{\nu J M}^{\nu' J' M'} \pm \hbar 
  \sqrt{(g_{\nu J M}^{\nu' J' M'})^2 f({\bf R})^2 + 
    (\delta_{\nu J M}^{\nu' J' M'})^2/4}\; , 
\end{eqnarray}
where $\bar E_{\nu J M}^{\nu' J' M'}$, $g_{\nu J M}^{\nu' J' M'}$, and
$\delta_{\nu J M}^{\nu' J' M'}$ are the parameters of the transition to
the strongest coupled upper rovibronic state $\ket{e \nu' J' M'}$. The sign is
determined by the sign of the detuning. During the passage the molecules
acquire transverse velocity due to the nonresonant dipole force, given by the
gradient of the energy Eq.~(\ref{eigenenergies}). We predict the occurrence of
different trajectories labelled by the quantum numbers of the relevant
rovibrational states. The quantum state dependent deflection angles
read
\begin{equation}
  \label{angle}
  \alpha(\nu J M)=\frac{v_z}{v_x} \approx \frac{v_{\mbox{\tiny rec}}\, (g_{\nu
  J M}^{\nu' J' M'})^2 \, l}{v_x^2 \, \delta_{\nu J M}^{\nu' J' M'}} \; ,
\end{equation}
where $l=\sqrt{A}$ is the effective interaction length, $v_{\mbox{\tiny
    rec}}=\hbar k/M$ is the photon recoil velocity.

In the following, we analyze the complex deflection pattern on the example of
the Na$_2$. The ground electronic state $X^1\Sigma^+_g$ is
coupled to the state $A^1\Sigma^+_u$ by a dipole-allowed optical transition
with $E_{\mbox{\tiny el}}=17541$ cm$^{-1}$ \cite{smirnov96}.
Both states are singlet; hence the electron spin plays no role
($\Sigma=\Sigma'=0$). Since $\Lambda=\Lambda'=0$, the selection rules $J'=J\pm
1$, $M'=M$ apply (the rotational Q branch is missing) and the H\"onl--London
factor is $S(J,J')=J+1$. According to theoretical calculations
\cite{stevens77}, confirmed by experiments \cite{ducas76}, the electric dipole
is $|{\bf d}_f^e|^2=14$ in atomic units.

\begin{figure}
\resizebox{\textwidth}{!}{%
        \includegraphics*{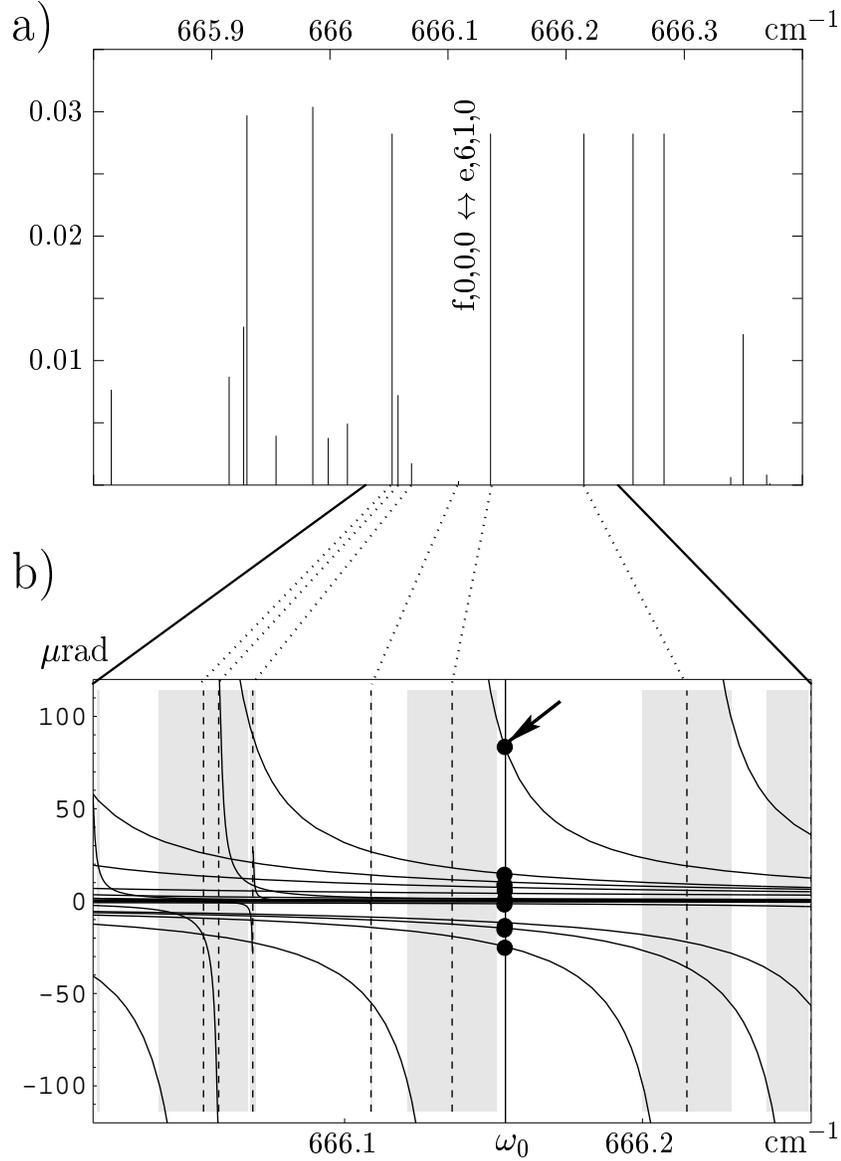}
}
\caption{ a) The 
      spectrum of Na$_2$ around the transition $f$, $\nu$=0, $J$=0,
      $M$=0 $\leftrightarrow e$, $\nu'$=6, $J'$=1,
      $M'$=0. $E_{\mbox{\tiny el}}$ is subtracted from the
      frequencies.  b) Deflection angles as a function of the laser
      frequency. Dashed vertical lines correspond to the
      transitions. Outside the grey regions $\delta_{\nu J M}^{\nu' J'
      M'} > 10 \; g_{\nu J M}^{\nu' J' M'}$ thus the non-resonance
      condition is fulfilled. At the specific laser tuning $\omega_0$
      several deflection angles occur (points). The arrow indicates
      the deflection of molecules in the vibrational-rotational ground
      state.}  \label{fig:angles}
\end{figure}

Suppose the laser is tuned close to resonance with the rovibronic
transition $f$, $\nu=0$, $J=0$, $M=0$ $\leftrightarrow$ $e$, $\nu'=6$,
$J'=1$, $M'=0$ (with energy $E_{\mbox{\tiny el}}+666.136$ cm$^{-1}$,
and natural linewidth $\Gamma= 3.4 \times 10^{-5}$ cm$^{-1}$).  The $0
\leftrightarrow 6$ vibrational transition is associated with one of
the largest Franck--Condon factors ($|R_0^6|^2=0.113$).  The
neighboring transitions are represented in Fig.~2a. For a temperature of
1000 K, the distribution can be truncated at 10 vibrational
and 100 rotational levels.  As the coupling strength is much less than
$B_\nu$, resonances that are as far as the rotational line separation
($B_\nu = 0.155$ cm$^{-1}$) can be safely neglected. Thus a molecule
has a chance to be deflected if its rovibrational state is involved in
the transitions shown in Fig.~2a. The adjacent lines are separated by
at least $0.05$ cm$^{-1}$; hence $g \approx 1.5 \times 10^{-3}$
cm$^{-1}$ can be chosen ($\Gamma \ll g \ll B_\nu$). Focusing the
standing wave to $l \approx 50 \mu$m, the necessary field intensity is
$I \approx 0.3$ mW.  The Raman--Nath regime implies a purely
geometrical condition on the deflection angle $\alpha$ that is limited
by $\alpha<\lambda/l\approx 10^{-2}$ radian. The velocity is set at
$v_0 = 500$~m/s, yielding an interaction time of 100 ns. The
probability of a spontaneous emission event is less than 10 \%; hence
neglecting the scattering force is justified. Deflection angles
according to the above parameters are shown in Fig.~2b as a function
of the laser frequency. Each curve belongs to a single
rotational-vibrational state.  The resonant region around each
transition is darkened.  Outside these regions the detuning is one
order of magnitude larger than the corresponding coupling constant,
thus the nonresonance condition is fulfilled.  For the laser frequency
$\omega_0$ at the solid vertical line, the detuning will be $\delta
\approx 0.02$ cm$^{-1}$.  The points represent the observable
deflection angles. The arrow indicates the angle corresponding to the
rotational-vibrational ground state (approximately 100 $\mu$rad; for
comparison, a single photon recoil would cause 35 $\mu$rad).
According to Fig.~2b, several deflected partial beams are present
simultaneously, which could be resolved depending on the beam
divergence and on the detection technique. We note that a deflected
beam does not correspond to a single quantum state, as we did not
include in our model all degrees of freedom of a molecule, e.g. the
nuclear spin was neglected. Thus the deflected beam contains all
possible nuclear spins, altogether 6 states.

The center-of-mass motion has been described in terms of classical
trajectories in our model, which offered an intuitive picture with
focus on the more essential internal part of the dynamics. We avoided,
however, some details that would degrade the predicted signal in an
experimental realization. There are several mechanisms giving rise to
a broadening of the deflected partial beams. First of all, we assumed
a well-defined longitudinal velocity $v_x$. It follows from
(\ref{angle}) that the double the relative variance in $v_x$
equals the relative variance of the deflection angle. Still we can
safely tolerate a few percent uncertainty which can easily be
achieved, even without velocity selection, by using supersonic beams.
Next, the molecule has to cross the field in a region where the
interaction strength, i.e. $f(\mbox{\bf R})$, is approximately
uniform. The deflection angle being proportional to $\sin{ 2 k z}$,
the crucial part in this respect is the direction $z$. The variation
$2 k {\Delta}z = \pm 0.45$ around $2 k z= \pi/2$ amounts to 10 \%
variation in the final deflection angle. However, the confinement of
the beam to a size smaller than the wavelength causes diffraction. The
transverse velocity that was supposed to be $v_z=0$ broadens to a
distribution with a width of $\hbar/(2 M {\Delta}z) = v_{\mbox{\tiny
rec}}/(2 k{\Delta}z)\approx v_{\mbox{\tiny rec}}$. Deviations in the
direction $y$ from the center line $y=0$ do not induce such drastic
diffraction since the gaussian beam waist is several orders of
magnitude larger than the wavelength. Therefore $v_y
\approx 0$ remains a good approximation.
Finally, the spontaneous photon emission is accompanied by a recoil
that introduces a randomness in the deflection angle. When the
spontaneous emission probability is kept below 10 \%, the occurence of
several photon recoil kicks can practically be excluded. Similarly to
the diffraction problem in the $z$ direction, the single photon recoil
raises the question of resolution. That is, the deflection broadening,
at most $v_{\mbox{\tiny rec}}/v_x$, must be dominated by the
deflection angle given in (\ref{angle}). Note that these two
quantities were 35 $\mu$rad and 100 $\mu$rad in our numerical example.

The effect of spontaneous emission is, however, a bit more complex
than just broadening the velocity distribution. Following a
spontaneous jump, the molecule exits the two states of the
quasiresonant transition and the process of accumulating transverse
momentum is interrupted. This gives rise to a weak background between
the zero and maximum deflection angles. The suppression of this
background to the possible minimum is the guiding principle when
chosing the appropriate detunings. In fact, this imposes the ultimate
condition on the generalization of the proposed scheme to other
molecules. The basic ingredient of our scheme is the ``isolated line''
in the spectrum. That is the spectral density must leave enough space
to tune the laser quasiresonant with a single transition without
introducing photon scattering. Roughly, an order of magnitude between
the detuning $\delta_{\nu J M}^{\nu' J' M'}$ and the linewidth
$\Gamma_{\nu J M}^{\nu' J' M'}$, and also an order of magnitude
between $\delta_{\nu J M}^{\nu' J' M'}$ and the distance to the
adjacent transitions in the spectrum is necessary. In Na$_2$, being a
light diatomic molecule, we had three orders of magnitude difference
between $\Gamma$ and the relevant transition line separations. This
suggests that the non-resonance condition can also be fulfilled for
heavier diatomic or polyatomic molecules as long as the rotational
constant $B_\nu$ is not less than one tenth of that of the sodium we
considered here. Decisive conclusions on the applicability cannot be
drawn without the detailed knowledge of the spectrum. Nevertheless, it
may always be possible to find accidentally well isolated lines
associated with the ground rovibrational state of the molecule.

In conclusion, we have shown that it is possible to deflect diatomic
molecules quantum state selectively. The very high sensitivity
originates from the monochromatic laser field that can be finely tuned
within the rovibrational band. The laser-induced coupling of the
vibrational and rotational modes to the center-of-mass translation
results in quantum correlations between the different motional degrees
of freedom. No analogue of this type of correlations exists in atom
optics. There only the quantized light mode (quantum prism
\cite{prizma}) or the internal electronic state (optical
Stern--Gerlach effect \cite{sleator}) can be entangled to the
translational motion. By contrast to atoms, however, in a usual beam
of molecules various rovibrational states are thermally populated;
hence the possibility of selecting the rovibrationally cold molecules
has important practical consequences. This arrangement may be
integrated in the preparation stage of experiments performing coherent
manipulation with molecular beams. The process still cannot be called
cooling, since thermal equilibrium is not maintained throughout. In
fact, it is a laser Maxwell's demon \cite{letokhov}, which is not in
contradiction to the second law of thermodynamics if the momentum
exchange with the laser is correctly taken into account
\cite{milburn}. The Maxwell's demon for molecules, beyond its unusual
statistical properties, would be of great practical value for preparing a
sample of non-vibrating and non-rotating molecules.

We are grateful for useful discussions with J\"org Schmiedmayer and Viktor
Szalay. This work was supported by the National Scientific Fund of Hungary
(OTKA) under contracts No. T023777, F032341, and F032346. T. K. acknowledges
the support of the Hungarian Academy of Sciences (Bolyai J\'anos Kutat\'asi
\"Oszt\"ond\'{\i}j).

\end{document}